\documentclass[twocolumn,amsmath,amssymb,aps,prb]{revtex4-2}
\usepackage{graphicx}
\usepackage{hyperref}
\usepackage{xcolor}
\usepackage{multirow}
\begin{document}

\preprint{APS/123-QED}

\title{Magnus spin Hall and spin Nernst effects in gapped 2D Rashba systems} 

\author{Priyadarshini Kapri$^{1,2}$}
\author{Bashab Dey$^{1}$}
\author{Tarun Kanti Ghosh$^{1}$}

\affiliation{$^{1}$Department of Physics, Indian Institute of Technology-Kanpur, 
Kanpur-208 016, India \\
$^{2}$Department of Physics, Osaka University, Osaka-560 0043, Japan}

\date{\today}

\begin{abstract}

We study the Magnus transport in a  gapped 2D electron gas with Rashba spin-orbit coupling using semiclassical Boltzmann transport formalism. 
 Apart from its signature in the charge transport coefficients, the inclusion of Magnus velocity in the spin current operator enables us to study Magnus spin transport in the system.
In particular, we study the roles of mass gap
 and Fermi surface topology on the behavior of Magnus Hall and Nernst conductivities and their spin counterparts. We find that the Magnus 
 spin Hall conductivity vanishes in the limit of zero gap, unlike the universal spin
   Hall conductivity $\sigma_s=e/(8\pi)$. The
    Magnus spin currents with spin polarization perpendicular to the applied bias (electrical/thermal) are finite while with polarization along the bias vanishes.
Each Magnus conductivity displays a plateau as Fermi energy sweeps through the gap and has peaks (whose magnitudes decrease with the gap) when the Fermi energy is at the gap edges. 

\end{abstract}

\maketitle

\section{Introduction}

After the discovery of classical Hall effect, a wide range of Hall effects (HEs),
 such as quantum HE, anomalous HE, spin HE, nonlinear HE and  thermal HE have been
discovered in the course of time. These discoveries have played a leading
 role in elucidating the novel electronic states \cite{Klitzing,Haldane} and
 electron dynamics \cite{Hall,Karplus} and thus have drawn tremendous interest to 
the solid state community. The traditional (classical) Hall \cite{Hall}
 effect and quantum Hall effect \cite{Klitzing} arise 
only in the presence of an
external magnetic field. However, anomalous \cite{Hall2,Sinitsyn,Nagaosa} and the quantum anomalous Hall
effect \cite{Chang,Liu,He} originate from the anomalous velocity of the charge
 carriers, where the role of Berry curvature (BC) is involved. Further,
 the involvement of Berry curvature  with spin index, valley index, and orbital degrees of freedom
lead to the spin HE \cite{Sinova1,Sinova2}, valley HE \cite{Xiao1,Xiao2,Mak} 
and orbital HE \cite{Go,Canonico} respectively. 
Furthermore, the dipole moment of BC  in momentum space  generates a net anomalous 
velocity and  gives rise to the non linear Hall effect \cite{Sodemann}. Moreover, the thermal
analogue of these Hall effects lead to thermal Hall 
effects \cite{Yu,Liang}, which play a pivotal role in the field of caloritronics.

As we know, the discrete symmetries of the Hamiltonian namely, inversion
symmetry (IS) and time reversal symmetry (TRS) play
a significant role in directing the fate of the Hall current.  For example, 
the intrinsic anomalous Hall effect (AHE) vanishes for TRS invariant systems, because the Berry curvature
 of the systems follows the relation: $\Omega({\bf k}) = -\Omega(-{\bf k})$ and hence the Hall conductivity (which is proportional to the integral of Berry curvature) vanishes. However, such a distribution of Berry curvature in a TRS invariant system raises the question whether 
it can give rise to any interesting phenomena in
charge transport. This leads to realization of the valley Hall effect \cite{Xiao1} and the
non-linear Hall effects in TRS invariant but
IS broken systems \cite{Sodemann,Liang,Morimoto,Nandy}. Very
recently, it has been observed that such TRS invariant but IS broken systems with a built-in electric field and no external magnetic field
manifests a new type of Hall effect namely,
Magnus Hall effect (MHE) \cite{Papaj}. 
The Magnus Hall effect represents the transverse force acting on a Bloch wave packet (having finite Berry curvature) moving under a potential gradient.	To have the Magnus `force' on an electron, it must have a finite Berry curvature, irrespective of the presence or absence of TRS.

In Refs. \cite{Mandal,Nag} the appearance of
Magnus Nernst effect (MNE) (transverse Magnus current produced by longitudinal thermal gradient) and Magnus thermal Hall
effect (Magnus heat current in a direction transverse to a temperature gradient) are demonstrated in TRS invariant but inversion
broken systems.
It has been proposed that the systems having IS breaking and TRS invariant nature, such as, monolayer (ML) graphene on
hBN, bilayer (BL) graphene with applied perpendicular
electric field \cite{McCann,Rozhkov}, the two-dimensional (2D)
transition metal dichalcogenides $\mathrm{MX_2}$ ($\mathrm{M=Mo,W}$  and
$\mathrm{X=S, Se, Te}$) \cite{Qian,Xiao1,You}, hetero-structures \cite{Yankowitz}, surfaces
of topological insulator (TI) \cite{Fu} and Weyl semimetals \cite{Hasan,Armitage,Yan} are the potential candidates for
 exploring the MHE and MNE.  Thus, it remains to be seen whether the Magnus Hall effect and Magnus Nernst effect exist in
 systems where both the IS and TRS are broken with non zero Berry curvature. 

Recently, in our previous paper \cite{Kapri},
we have shown that the origin of the spin Hall current 
 can be explained by redefining the spin
current operator with inclusion of the anomalous velocity. We have also 
shown that the anomalous velocity can generate pure
 anomalous non linear spin current with in-plane polarization.
 Thus, our general instinct
 motivates us to investigate whether the inclusion of Magnus velocity in spin
 velocity operator can produce Magnus spin Hall current (out of plane polarization)
 and Magnus spin current (in-plane polarization). Also, the observation of
 Spin Nernst effect \cite{Meyer,Sheng} 
 points towards a possibility of  Magnus spin Nernst effect (MNE) on incorporation of the Magnus velocity.

Motivated by the above discussions, we study the Magnus Hall and the Magnus 
Nernst  effect of a 2D Rashba system with
 mass gap that causes the TRS breaking of the system. In particular,
 we investigate the role of gap parameter (induced by the TRS breaking 
 gap term) and the role of Fermi surface topology in the behavior of
 Magnus Hall and the Magnus Nernst conductivities. Furthermore, we redefine 
the spin current operator with including the Magnus velocity and study 
the spin counterparts of Magnus conductivities. Our studies mainly emphasize on the results obtained using the conventional definition (with  Magnus velocity) of spin current operator. A modified spin current operator was introduced in Ref.\cite{Shi} which has a term representing spin torque in addition to the conventional terms of the spin current operator. We have also commented on the contribution of this additional term in the Magnus transport coefficients. It is to be noted that a lot of debates ensued regarding the definition of spin current operators \cite{Rashba2,Rashba3,Shi,Sun1,Wang1,Wang2}. Later, it was shown that in a spin non-conserving system, the equilibrium spin current and violation of the Onsager relation are intrinsic properties of spin transport irrespective of the definition of spin current operator \cite{Sun2}.


This paper is organized as follows. In Sec. \ref{sec2A}, we present the semiclassical Boltzmann
transport formalism to discuss the Magnus Hall conductivities (MHCs) 
in the diffusive regime  as well as in the ballistic regime. Section \ref{sec2B} presents the 
extension of above theory to study spin counterparts of MHCs by 
inclusion of Magnus velocity in conventional spin velocity operator. Section \ref{sec3} includes the basic
information of a 2D gapped Rashba system. In Sec. \ref{sec4}, we
present the analytical and numerical results. Finally Sec. \ref{sec5} concludes and summarizes
our main findings.

\section{Formalism of Magnus transport}
\label{seca}
\begin{figure}
\begin{center}
\includegraphics[width=85mm,height=57mm]{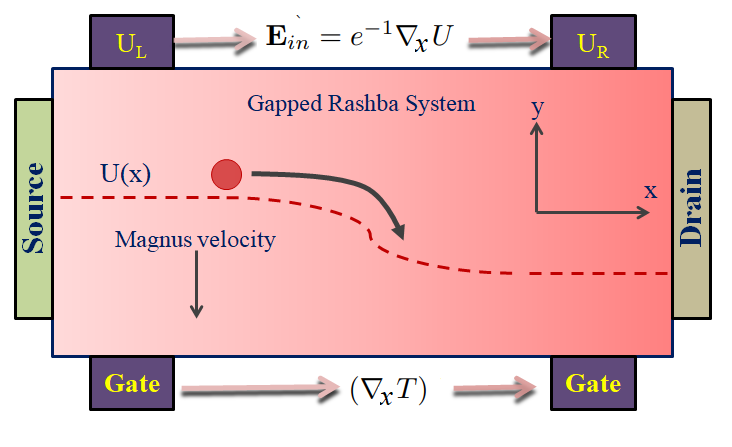}
\caption {Schematic illustration of the device for Magnus Hall effect. The difference in 
potential energies (induced by two gates) gives rise to an in-built electric field ${\bf E}_{in} = e^{-1} {\bf\nabla}_x U$ along the bar.
 In presence of the in-built electric field, a self-rotating electron wave-packet
(with a finite BC) that exits the
source manifests a Magnus shift in transverse direction.  An additional  electric field ${\bf E}$ , or temperature gradient
(${\bf \nabla_x} T $)  is applied between the source and drain. }
\label{Fig1}
\end{center}
\end{figure} 
In this section, we present a theory for
Magnus Hall, Magnus Nernst  conductivities for a generic two-band system  and finally extend the theory to calculate their spin counterparts.

\subsection{Derivation of Magnus transport coefficients}
\label{sec2A} A mesoscopic Hall bar is considered, where
a slowly varying electric potential
energy $U(r)$ is set along the length of the bar (See Fig. \ref{Fig1}) 
The potential energy gradient is introduced
by the two gate voltages, $U_L$ and $U_R$. 
The difference in potential energies ($\Delta U = U_L - U_R$) gives rise to an in-built
 electric field ${\bf E}_{in} = e^{-1} {\bf\nabla}_x U$ ($-e$ being the electronic charge) along the bar. Furthermore, 
 	an additional electric field ${\bf E}$ (small bias), or temperature gradient
 	(${\bf \nabla_x} T $) is applied between the source and drain.
 

As mentioned earlier, to have the Magnus transport, the material of the
bar should have a non-zero Berry curvature ${\bf \Omega}$.
The motion of wave-packet inside the Hall bar is described
 by the semiclassical equations of motion \cite{Xiao,Son}
\begin{eqnarray}
\label{eq1}
	\hbar \dot{{\bf r}}&=&\nabla_{{\bf k}}\epsilon_{{\bf k}}+(\nabla_{\bf r} U +e{\bf E})\times{\bf \Omega},\\
\hbar \dot{{\bf k}}&=-&\nabla_{\bf r} U-e{\bf E}.
\end{eqnarray}
The first, second and third terms in the right hand side of
Eq. (\ref{eq1}) are associated with the semiclassical band 
velocity ${\bf v}_b=\frac{1}{\hbar}\nabla_{\bf k}\epsilon_{\bf k}$,
  Magnus velocity ${\bf v}_{m} = \frac{1}{\hbar}(\nabla_{\bf r} U \times {\bf\Omega}$) and anomalous
 velocity ${\bf v}_{a}$=$\frac{e}{\hbar}({\bf E} \times {\bf \Omega}$), 
respectively with ${\bf \Omega}$ being the Berry curvature. For a 2D system confined in 
$x$-$y$ plane, the Berry curvature is always in the $\hat{z}$-direction.   Hence both the Magnus
and the anomalous velocity are along the $\hat{y}$-direction  because  the in-built potential gradient and  the external bias (electric field or temperature gradient) are along the $\hat{x}$ direction.

Within the relaxation time approximation, the non-equilibrium carrier distribution function
$f ({\bf r}, {\bf k})$ obeys the Boltzmann transport 
equation (BTE) \cite{Ashcroft}
\begin{equation}
\label{eq3}
\frac{\partial f}{\partial t}+\dot{{\bf r}}\cdot\frac{\partial f}{\partial {\bf r}}+\dot{{\bf k}}\cdot\frac{\partial f}{\partial {\bf k}}=-\frac{f-f_0}{\tau},
\end{equation}
where $f_0$ and $\tau$ denote the 
 distribution function (in the presence
of an inbuilt electric field) with no external bias and 
the scattering time, respectively. For simplicity, $\tau$
is considered to be momentum independent. In the steady state condition 
($\frac{\partial f}{\partial t} = 0$), with no external bias 
 the
 distribution function is given by the Fermi
function
\begin{equation}
f_0 ({\bf k}, {\bf r}) =\frac{1}{1+e^{\beta[\epsilon({\bf k},{\bf r})-\epsilon_F]}},
\end{equation}
where $\epsilon({\bf k}, {\bf r}) = \epsilon_{\bf k} + U ({\bf r})$, $\beta=1/k_BT$
and $\epsilon_F$ is a constant Fermi energy.   Now, in the
presence of an external field ($\hat{x}$- directed), the
 distribution function up to
first order in the bias fields can be obtained as
\begin{equation}
\label{eq5}
f = f_0 + v_{b,x}\tau (e  E_x + \beta [\epsilon({\bf k}, {\bf r}) -\epsilon_F] k_B \nabla_x T) \frac{\partial f_0}{\partial \epsilon_{\bf k}},
\end{equation}
where $v_{b,x}$ denotes the $x$ component of band velocity $v_b$. For our setup,  $v_{x}=v_{b,x}$,
 as both the anomalous and Magnus velocities are in $\hat{y}$ direction.

In the presence of external electric field ${\bf E}$ and
temperature gradient ${\boldsymbol{ \nabla}} T$ applied between the source and
drain, the charge current density ${\bf J}$ 
from linear response theory can be written as
\begin{eqnarray}
J_i = \sigma_{ij} E_j + \alpha_{ij} (-\nabla_j T ),\\\nonumber
\end{eqnarray}
where $i$ and $j$ denote the propagation and applied electric field directions with
$\sigma$  and $\alpha$ 
being the conductivity tensors. For our setup, $j=x$, as the external electric 
field or the temperature gradient is applied in $\hat{x}$ direction.

For a two-band system, the  general expression of charge current (considering all orders) in a 2D system 
can be written as 
\begin{equation}
{\bf J} =-e\sum_{\lambda,n}\int\frac{d^2{\bf k}}{(2\pi)^2} \langle\lambda,{\bf k| }\hat{v}|\lambda,{\bf k}\rangle f_n,
\label{eq7}
\end{equation}
where $\lambda$ denotes the band index, $f_n$ denotes the $n$-th  order ($n$: order of extrinsic bias field)  distribution function in the bias field and  $\hat{v}=\hat{v}_b+\hat{v}_a+\hat{v}_m$ with $\hat{v}_b$, $\hat{v}_a$ and $\hat{v}_m$ being the band,
anomalous and Magnus velocity operators.  In Eq. (\ref{eq5}), the first and second  terms denote the distribution function with no external bias ($f_0$) and the first order  distribution function  in bias fields ($f_1$), respectively.
From Eq. (\ref{eq7}), one can easily separate out the band velocity ($\hat{v}_b$),
 anomalous velocity ($\hat{v}_a$) and the Magnus velocity ($\hat{v}_m$) contributions in the
transport coefficients. Since the main focus of our paper is Magnus transport,
we discuss the Magnus contribution only. 


For an electric or thermal bias along $x$-direction, the Magnus Hall and Magnus Nernst currents of the first-order in bias field are given by $J_m=\sigma_m E_x$ and $J_m=-\alpha_m\nabla_x T$ respectively. The Magnus Hall ($\sigma_m$) and Magnus Nernst ($\alpha_m$) conductivities in the diffusive limit can be written as
\begin{eqnarray}
\sigma_m &=&-\frac{e^2\Delta U}{\hbar L}\sum_{\lambda}\int \frac{d^2{\bf k}}{(2\pi)^2}\Omega_z^{\lambda}v^{\lambda}_x\tau_{\lambda}\frac{\partial f_0}{\partial \epsilon_{\bf k}},\\
\alpha_m &=&\frac{e\Delta U}{\hbar L T}\sum_{\lambda}\int\frac{d^2{\bf k}}{(2\pi)^2}\Omega_z^{\lambda}v^{\lambda}_x\tau_{\lambda}\epsilon^{\prime}_{\bf k}\frac{\partial f_0}{\partial \epsilon_{\bf k}},
\end{eqnarray}
where  $\langle\lambda,{\bf k| }\hat{v_m}|\lambda,{\bf k| }\rangle=\frac{\partial U}{\partial x}\frac{\Omega_z^\lambda}{\hbar}$ and $f_1= eE_x v_x^{\lambda}\tau_{\lambda}\frac{\partial f_0}{\partial \epsilon_{\bf k}}$ (driven by electric field)  or $ \frac{1}{T} v_x^{\lambda}\tau_{\lambda}\epsilon^{\prime}_{\bf k}(\nabla_x T)\frac{\partial f_0}{\partial \epsilon_{\bf k}}$ (driven by temperature  gradient, see Eq. {\ref{eq5}})),  $U$ is a slowly varying function of $x$, $\partial U/\partial x =
\Delta U/L$ with $L$ being the length of the Hall bar 
 and $\epsilon_{\bf k}^{\prime}=(\epsilon_{\bf k}-\epsilon_F)$.

Here, it is to be noted that the  Magnus conductivities of linear order (in external bias fields) arise from $f_1$,
 whereas $f_0$ is responsible for first order anomalous conductivities as ${\bf v}_a \propto {\bf E}$. The Magnus conductivities can be viewed as an effective
second order response as external bias and the built-in electric field 
both are involved in the calculation of currents.


So far the Magnus responses in the diffusive limit has been discussed.  In the ideal ballistic regime, the mean free time between
two collisions is infinite $\tau \rightarrow \infty$. In a realistic setup, the ballistic limit corresponds to the case where the mean free path is much larger than the system length i.e. $v_x \tau\gg L$. So, no collision occurs in the transport direction along the Hall bar. Therefore, the right
hand side of the Boltzmann transport equation given in
Eq. (\ref{eq3}) vanishes in the ballistic regime. 


In this setup,  the larger electrochemical potential of the source region causes a surplus of electrons
entering the system at $x = 0$ interface with $v_{x}>0$. These electrons propagate across the device without any
scattering in the ballistic limit. Hence, only the carriers from the source with
positive velocity are allowed in region $0 < x < L$. Then, the
the distribution function can be
written as \cite{Papaj}
\begin{eqnarray}
f_1({\bf k},{\bf r})&=&-\Delta
\epsilon_F\frac{\partial f_0}{\partial \epsilon_{\bf k}}-\frac{\epsilon^{\prime}_{\bf k}}{T}\Delta T\frac{\partial f_0}{\partial \epsilon_{\bf k}} \hspace{0.1in} \mathrm{for} \hspace{0.1in} v_{x} >0,\\\nonumber
f_1({\bf k},{\bf r})&=&0\hspace{0.1in} \mathrm{for} \hspace{0.1in} v_{x} <0,
\end{eqnarray}
where $\Delta\epsilon_F=eV_{sd}$ is the electrochemical potential difference between
 the source and drain with $V_{sd}$ being the small bias voltage.
These solutions become identical with the solutions in Eq. (\ref{eq5}),
if one identifies the scattering length $v_x \tau =L$,
 $\Delta\epsilon_F/L= -e E_x$  and $\Delta T/L=-\nabla_x T$.

Thus, the linear order (in the bias field) Magnus Hall conductivity and the Magnus Nernst conductivity  in the ballistic regime  can be expressed as \cite{Papaj,Mandal}
\begin{eqnarray}
\label{eq17}
\sigma_m &=&-\frac{e^2\Delta U}{\hbar }\sum_{\lambda}\int_{{v}_x^{\lambda} >0} \frac{d^2{\bf k}}{(2\pi)^2}\Omega_z^{\lambda}\frac{\partial f_0}{\partial \epsilon_{\bf k}},\label{eq11}\\
\alpha_m &=&\frac{e}{\hbar T}\Delta U\sum_{\lambda}\int_{{v}_x^{\lambda} >0}\frac{d^2{\bf k}}{(2\pi)^2}\Omega_z^{\lambda}\epsilon^{\prime}_{\bf k}\frac{\partial f_0}{\partial \epsilon_{\bf k}}\label{eq12}.
\end{eqnarray}
As mentioned earlier, the
Magnus responses are dependent on the built-in electric
field.

In the following discussion, we elaborate the difference between anomalous and Magnus contributions. Under an inbuilt potential gradient ($\partial U/\partial x$) and applied source-drain bias ($E_x$) along $x$ direction, the transverse charge current in the ballistic limit upto linear order in $E_x$ is given by
	\begin{eqnarray}
		\label{eq13}
	J_y&=&\frac{e}{(2\pi)^2\hbar}\sum_{\lambda}\bigg[-\int d^2{\bf k}~f_0 \frac{\partial \epsilon_{\bf k}}{\partial k_y} + \int d^2{\bf k}~f_0  \frac{\partial U}{\partial x} \Omega_z ({\bf k}) \\\nonumber
	&+&\int d^2{\bf k}~f_0  e E_x \Omega_z ({\bf k})
	-\int_{v_x>0} d^2{\bf k}~e E_x L \frac{\partial f_0}{\partial \epsilon_{\bf k}}\frac{\partial \epsilon_{\bf k}}{\partial k_y} \\\nonumber
	& +& \int_{v_x>0} d^2{\bf k}~e E_x L \frac{\partial f_0}{\partial \epsilon_{\bf k}} \frac{\partial U}{\partial x} \Omega_z ({\bf k})\bigg]
	\end{eqnarray}
	The above expression contains five terms which are explained as follows: 
		(i) The first term, representing {\it equilibrium current}  due to band velocity, vanishes for every system. 
		(ii) The second term denotes an {\it equilibrium Magnus current}, as it appears in the system without any bias voltage but only due to an intrinsic potential gradient.
	 (iii) The third term represents the usual {\it anomalous Hall current} in response to the bias.
	In TRS invariant systems, the Berry curvature is an odd function in momentum space due to which its total over the entire Brillouin zone vanishes. As a result, a TRS invariant system produces no {\it equilibrium Magnus} or {\it anomalous Hall current}. However, for TRS broken systems, the {\it equilibrium Magnus} and {\it anomalous Hall currents} may be non-zero. 
	(iv) The fourth term arises from the Fermi surface anisotropy and depends on its orientation relative to the direction of applied bias. 
		 (v) The fifth term represents the {\it Magnus Hall current} (equivalent to Eq. (\ref{eq11})). Since the integration is only over states with positive $v_x$, the {\it Magnus Hall current} may exist in TRS invariant systems under certain conditions i.e. tilt or anisotropy in the dispersion. In a TRS broken system with even function of Berry curvature such as the gapped Rashba system, the {\it Magnus Hall current} will always be non-zero. Thus, it may be finite in both TRS invariant and TRS broken systems. Since the TRS invariant systems do not have {\it equilibrium Magnus} or {\it anomalous Hall} contributions, the Magnus Hall current is of special interest in those cases. However, going by the above discussion, the TRS should not be a criteria to define the {\it Magnus Hall current}. 


\subsection{Derivation of Magnus spin conductivities}
\label{sec2B}
Now we extend the above theory to calculate the spin counterparts of the Magnus conductivities. 
The general definition of spin current (considering all orders in bias fields) for a 2D system can be written as
\begin{equation}
\mathcal{J}_{ij}=\frac{\hbar}{2}\sum_{\lambda,n}\int\frac{d^2{\bf k}}{(2\pi)^2}\langle\lambda,{\bf k}|\hat{v}_{ij}|\lambda,{\bf k}\rangle f_n,
\end{equation}
where $\hat{v}_{ij}=\hat{v}_{b,ij}+\hat{v}_{a,ij}+\hat{v}_{m,ij}$ with $\hat{v}_{b,ij}$, $\hat{v}_{a,ij}$, and $\hat{v}_{m,ij}$ being the 
band, the anomalous and the Magnus components of 
spin velocity operator, respectively. The first index $i$ denotes the propagation direction and second index $j$ denotes the polarization direction.  Here, it is to be noted that  the conventional definition of spin current operator is given by
$\hat{{v}}_{b,ij} = (\hat{v}_{b,i} \sigma_{j} + 
\sigma_{j} \hat{v}_{b,i} )/2$,
 with 
$\hat{v}_{b,i} $  being  the 
	band velocity operator in $i$ direction. However, here the modified spin current operator contains both the anomalous  and Magnus velocity contributions. As mentioned earlier, our main focus is on the Magnus transport. Thus, by separating out the Magnus contribution, the Magnus spin current of first order (in the bias fields) can be written as
 \begin{equation}
\mathcal{J}_{m,ij}=\frac{\hbar}{2}\sum_{\lambda}\int\frac{d^2{\bf k}}{(2\pi)^2}\langle\lambda,{\bf k}|\hat{v}_{m,ij}|\lambda,{\bf k}\rangle f_1.
\label{eq14}
\end{equation}
with $f_1= eE_x v_x^{\lambda}\tau_{\lambda}\frac{\partial f_0}{\partial \epsilon_{\bf k}}$ (driven by electric field) or $\frac{1}{T} v_x^{\lambda}\tau_{\lambda}\epsilon^{\prime}_{\bf k}(\nabla_x T)\frac{\partial f_0}{\partial \epsilon_{\bf k}}$ (driven by temperature gradient, see Eq. {\ref{eq5}}).

From Eq. (\ref{eq14}), one can easily find out the Magnus spin conductivity. In the 
 diffusive regime, the $\hat{x}$ directed electric field (external) driven Magnus spin conductivity
 ($\mathcal{\sigma}_{m,yj}=\frac{\mathcal{J}_{m,yj}}{E_x}$)  of first order 
 with propagation and polarization in $\hat{y}$ and $\hat{j}$ directions, respectively,  can be written as
\begin{eqnarray}
\label{eq15}
\mathcal{\sigma}_{m,yj}&=&\frac{\hbar e}{2}\sum_{\lambda}\int\frac{d^2{\bf k}}{(2\pi)^2}\tau_{\lambda}v_x^{\lambda}v_{m,yj}^{\lambda} \frac{\partial f_0}{\partial \epsilon_{\bf k}},\\\nonumber
&=&\frac{e\Delta U}{2L}\sum_{\lambda}\int\frac{d^2{\bf k}}{(2\pi)^2}\tau_{\lambda}v_{x}^{\lambda}\sigma_{j}^{\lambda}\Omega_z^{\lambda}\frac{\partial f_0}{\partial \epsilon_{\bf k}},
\end{eqnarray}
where $v_{m,yj}^{\lambda}=\langle\lambda,{\bf k}|\hat{v}_{m,yj}|\lambda,{\bf k}\rangle=\frac{1}{2}\langle\lambda,{\bf k}|\hat{v}_{m,y}\sigma_j+\sigma_j\hat{v}_{m,y}|\lambda,{\bf k}\rangle=\frac{\Delta U\Omega_z^{\lambda}}{\hbar L}\sigma_{j}^{\lambda}$ with $\sigma_{j}^{\lambda}=\langle\lambda,{\bf k}|\sigma_j|\lambda,{\bf k}\rangle$.

Similarly, thermally driven first order Magnus spin conductivity ($\mathcal{\alpha}_{m,yj}=\frac{\mathcal{J}_{m,yj}}{-\nabla_x T}$)
 propagating in $\hat{y}$ direction with polarization in $\hat{j}$ direction can be  obtained as
\begin{eqnarray}
\label{eq16}
\alpha_{m,yj}
=-\frac{\Delta U}{2L T}\sum_{\lambda}\int\frac{d^2{\bf k}}{(2\pi)^2}\tau_{\lambda}v_x^{\lambda}\sigma_{j}^{\lambda} \Omega_z^{\lambda} \epsilon^{\prime}_{\bf k}\frac{\partial f_0}{\partial \epsilon_{\bf k}}.
\end{eqnarray} 
For the ballistic regime, the Eq. (\ref{eq15}) and Eq. (\ref{eq16}) becomes
\begin{eqnarray}
\label{eq17}
\sigma_{m,yj}
=\frac{e\Delta U}{2}\sum_{\lambda}\int_{v_x^{\lambda} >0}\frac{d^2{\bf k}}{(2\pi)^2}\sigma_{j}^{\lambda}\Omega_z^{\lambda} \frac{\partial f_0}{\partial \epsilon_{\bf k}},
\end{eqnarray}
\begin{eqnarray}
\label{eq18}
\alpha_{m,yj}
=-\frac{\Delta U}{2T}\sum_{\lambda}\int_{v_x^{\lambda} >0}\frac{d^2{\bf k}}{(2\pi)^2}\sigma_{j}^{\lambda} \Omega_z^{\lambda} \epsilon^{\prime}_{\bf k}\frac{\partial f_0}{\partial \epsilon_{\bf k}},
\end{eqnarray} 
where $v_x^{\lambda}\tau_{\lambda}$ is replaced by $L$. 
Here, it is worth noting that $\sigma_{m,yz}$ 
and $\alpha_{m,yz}$ denote the Magnus spin Hall conductivity and Magnus spin Nernst conductivity,
respectively (as the external bias field, propagation and polarization directions are mutually perpendicular to each other).


\section{2D gapped Rashba system}
\label{sec3}
Here we provide the basic information of a gapped two-dimensional electron gas (2DEG) system with the Rashba spin-orbit 
interaction (RSOI). The Hamiltonian of the system is given by
\begin{eqnarray}\label{ham-rash2d}
\label{eq10}
H&=&\frac{{ \hbar^2 \bf k^2}}{2m^{*}} \sigma_0 + 
\alpha \boldsymbol\sigma \cdot ({\bf k} \times {\bf \hat{z} }) + M \sigma_z,
\end{eqnarray} 
where $m^{*}$ denotes the effective mass of a charge carrier with ${\bf k}=\{k\cos\phi,k\sin\phi\}$ being the wavevector of the charge carrier,  
$\sigma_0$ is the $2 \times 2$ identity matrix,  
$\sigma_{x,y,z}$ are the Pauli's spin matrices and  $\alpha$ is the RSOI strength 
which is responsible for inversion symmetry (IS) breaking of the system.
The term $M\sigma_z$  breaks time reversal symmetry (TRS) of the system and can be created either by application of  an 
external magnetic field \cite{Culcer}
or by applying a circularly polarized electromagnetic radiation \cite{Ojanen}.

The energy spectrum of the system is obtained as
\begin{equation}
\label{eqd}
\epsilon_{\lambda}({\bf k}) = \frac{\hbar^2 k^2}{2m^{*}} + 
\lambda\sqrt{M^2+\alpha^2 k^2},
\end{equation}
with $\lambda=\pm$ denoting the band indices.
The corresponding normalized wavefunctions  are  
\begin{eqnarray}
\label{eqwf}
|+, {\bf k}\rangle  = 
\left[\begin{array}{c}
\cos\frac{\theta}{2}e^{-i\phi}   \\
-i\sin\frac{\theta}{2}
\end{array} \right]\hspace{0.02in};\hspace{0.02in}
 |-,{\bf k}\rangle  = 
\left[\begin{array}{c}
\sin\frac{\theta}{2}e^{-i\phi}   \\
i\cos\frac{\theta}{2}
\end{array} \right],
\end{eqnarray}
with $\cos\theta= M/\sqrt{M^2+\alpha^2 k^2}$ and 
$\sin\theta = \alpha k/\sqrt{M^2+\alpha^2 k^2}$. The TRS breaking term $M\sigma_z$ creates a finite gap $2M$ at $k=0$.
 The spin orientation of an electron for this system is obtained as
$\sigma^{\lambda}=\langle \lambda,{\bf k}|\boldsymbol\sigma| \lambda,{\bf k}\rangle= 
\lambda\{\sin\theta\sin\phi, -\sin\theta\cos\phi, \cos\theta\}$.

\begin{figure}
\includegraphics[width=89mm,height=60mm]{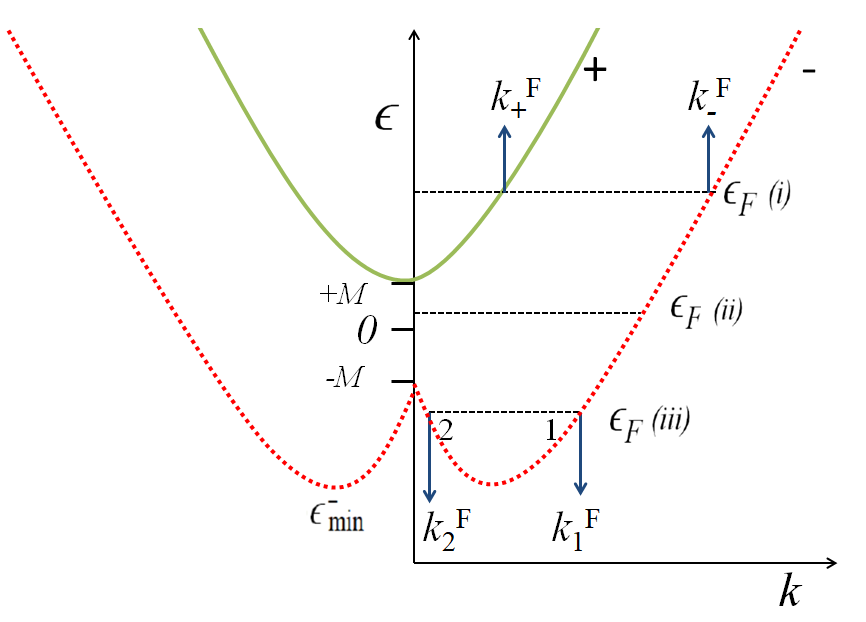}
\caption {Sketch of spin-split band structure of a  2D Rashba system with  $M\sigma_z$, when $M <2 \epsilon_{\alpha}$ with $\epsilon_{\alpha} = m^*\alpha^2/(2\hbar^2)$. 
The band $\epsilon_+({\bf k})$ has a minimum energy 
$\epsilon_{\rm min}^{+} = +M$ located at $k=0$ for all values of $M$, whereas
the band $\epsilon_-({\bf k})$ has a minimum energy 
$\epsilon_{\rm min}^{-} = -\epsilon_{\alpha}(1 + \tilde M^2) $  
at $k_m = k_\alpha \sqrt{1 - \tilde M^2}$.
It should be noted  that  expression for
$\epsilon_{\rm min}^{-}$ is valid only when $\tilde M < 1$.  }
\label{Fig2}
\end{figure}

The wavevectors corresponding to  regime $(i)$  ($\epsilon>M$, see Fig. \ref{Fig2}) are
$k_{\lambda}=k_{\alpha} \sqrt{ (\tilde{E}-\lambda)^2-\tilde{M}^2} $,
with $\tilde{E} = \sqrt{1 + \tilde{\epsilon} + \tilde{M}^2}$,
$\tilde{\epsilon} = \epsilon/\epsilon_{\alpha}$,  $k_\alpha = m^*\alpha/\hbar^2$.
The wavevectors $k_{\pm}$ represent the radii of the two concentric circular constant energy contours of convex shape. 
For regime $(iii)$ ($\epsilon< -M$ and $M<2\epsilon_{\alpha}$), the topology of the Fermi surface has convex-concave shape on
the outer and inner branches, respectively. Here, the wavevectors are presented by  $k_{\nu}=k_{\alpha} 
\sqrt{ [1+(-1)^{\nu-1}\tilde{E}]^2-\tilde{M}^2 }$, where $\nu=1,2$ ($\nu=1\rightarrow$ outer branch and $\nu=2\rightarrow$ inner branch).  
For the regime regime $(ii)$ ($-M\le\epsilon\le M$), only the
branch $\nu=1$ with $\lambda=-1$ exists.

For regime $(i)$, the velocity component $v_x$ corresponding to band $\lambda$ is obtained as 
${v}_{x}^{\lambda}=\frac{\hbar k_{\alpha}}{m^*}\tilde{E}\Big[1-\frac{\tilde{M}^2}{(\tilde{E}-\lambda)^2}\Big]^{1/2}\cos\phi$.  This
 yields the limit of  $\phi$ integration for calculating different Magnus conductivities in ballistic regime (see Eqs. (\ref{eq11}-\ref{eq12}) and Eqs. (\ref{eq17}-\ref{eq18})).
In the regime ($iii$),  $v_x$
can be obtained from the same equation with $\lambda  = -1$ 
and $\tilde{E}$ replaced by $(-1)^{\nu-1}\tilde{E}$.   
For the regime $(ii)$, 
$v_x$ has the similar form with $\nu=1$.

The IS and TRS breaking terms give rise to a non zero Berry curvature in the system. The Berry curvature  corresponding to band $\lambda$ is
given by
\begin{equation} 
{\boldsymbol \Omega}_\lambda({\bf k}) =  
-\lambda \frac{M  \alpha^2 {\bf \hat z} }
{2(M^2 + \alpha^2k^2)^{3/2} },
\end{equation}
which is isotropic in nature. It peaks at 
$ k = 0$ and decays with increasing $ k$.
For a gapless 2D Rashba system, the Berry curvature and the orbital magnetic moment (OMM) are zero.


\section{Results and Discussion}
\label{sec4}

This section  presents the results of different Magnus conductivities and their
 spin  counterparts along with corresponding discussions.
The plots for different conductivities  are presented  as a function of scaled
 Fermi  energy $\tilde{\epsilon}_F$  ($\tilde{\epsilon}_F=\epsilon_F/\epsilon_{\alpha}$)
for different values of dimension less gap parameter $\tilde{M}$ ($\tilde{M}=M/(2\epsilon_{\alpha})=0.05,0.10,0.15$).
The temperature for all the plots is fixed at $T=1$ K. 

\subsection{Electric field driven Magnus conductivity and Magnus spin conductivities}
\begin{figure}
\begin{center}
\includegraphics[width=80mm,height=60mm]{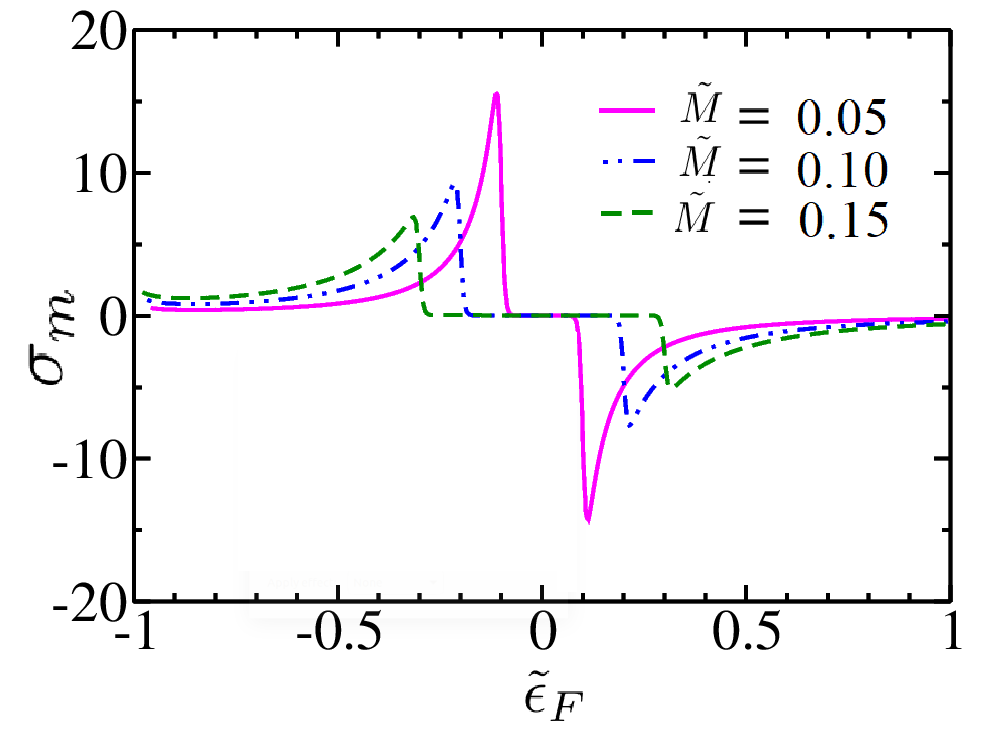}\caption {The Magnus Hall conductivity $\sigma_m$ (in units of 
$\mathcal{\sigma}_{0}$)
as a function of 
scaled Fermi energy  $\tilde \epsilon_F$ for  
different values of  $\tilde{M}$. Temperature is fixed at $T=1$ K.}
\label{FigMH}
\end{center}
\end{figure}  
We calculate the Magnus Hall conductivity using Eq. (\ref{eq11}). 
At zero temperature the Magnus Hall conductivity in ballistic regime has the following forms
\begin{eqnarray}
\label{eq23}
\sigma_m &=&-\mathcal{\sigma}_{0}\frac{4\tilde{M}}{(\tilde{\epsilon}_F+\tilde{M}^2)^2}
\hspace{0.05in} \mathrm{for \hspace{0.05in} regimes} \hspace{0.05in} (i), \\\nonumber
\sigma_m &=&\mathcal{\sigma}_{0}\frac{\tilde{M}}{\tilde{E}_F(\tilde{E}_F+1)^2}  \hspace{0.1in} \mathrm{for \hspace{0.05in} regime} \hspace{0.05in} (ii),\\\nonumber
\sigma_m &=&\mathcal{\sigma}_{0}\frac{2\tilde{M}(1+\tilde{E}_F^2)}{\tilde{E}_F(\tilde{\epsilon}_F+\tilde{M}^2)^2}
\hspace{0.05in} \mathrm{for \hspace{0.05in} regimes} \hspace{0.05in} (iii),
\end{eqnarray}
where $\tilde{E}_{F} = \sqrt{1 + \tilde{\epsilon}_{F} + \tilde{M}^2}$ with 
$\tilde{\epsilon}_{F} = \epsilon_F/\epsilon_{\alpha}$ and $\mathcal{\sigma}_{0}=\frac{e^2}{\hbar}\frac{\Delta U }{16\pi\epsilon_{\alpha}}$.

 It is to be noted that $\sigma_m$ is discontinuous at $\epsilon_F=\pm M$. The reason is discussed below.
The expression of $\sigma_m$ (see Eq. {\ref{eq11}}) can be re written as
 $\sigma_m =-\frac{e^2\Delta U}{\hbar (2\pi)^2 }\sum_{\lambda}
 \int_{-\infty}^{\infty}d\epsilon_k \mathcal{H}_{\lambda}(\epsilon_k)\frac{\partial f_0}{\partial \epsilon_{\bf k}}\approx-\frac{e^2\Delta U}{\hbar (2\pi)^2 }\sum_{\lambda}\mathcal{H}_{\lambda}(\epsilon_F) $, 
where $\mathcal{H}_{\lambda}(\epsilon)=\int_{{v}_x^{\lambda} >0}d\phi k_{\lambda}\frac{dk_{\lambda}}{d\epsilon_k} \Omega_z^{\lambda}$. 
Thus, the continuity of the Magnus Hall conductivity depends on 
the continuity of $\sum_{\lambda}\mathcal{H}_{\lambda}(\epsilon_F)$.
From Fig. \ref{Fig2}, 
it can be seen that `$+$' band and branch `$2$' of `$-$' band are discontinuous at $M$ and $-M$ respectively. Since $\mathcal{H}_{+}(\epsilon_F=M)\neq0$ and $\mathcal{H}_{-,2}(\epsilon_F=- M)\neq0$ , $\sigma_m$ becomes discontinuous at $\epsilon_F=\pm M$.
It is worth mentioning that the linear Hall and spin Hall currents in a 2D gapped Rashba system  are continuous at $\pm M$ \cite{Culcer,Kapri}, although
the integrands in their expressions are discontinuous at $\pm M$. This is because the linear Hall and spin Hall
currents arise from $f_0$ (as opposed to $f_1\propto\frac{\partial f_0}{\partial \epsilon}=-\delta(\epsilon-\epsilon_F)$  in Magnus Hall) 
due to which integration is performed from $-\infty$ to $\epsilon_F$ (at $T\to0$) and not just the value of integrand at $\epsilon_F$ is picked up, as was the case with Magnus Hall current. 
For a 2D gapped system, similar to Magnus Hall current, non linear 
 spin currents are also discontinuous at $\pm M$ because of the presence of $f_1$ in their definition.
 
It is worthwhile to mention that the asymmetric Berry curvature for right and left moving modes on the Fermi surface is necessary to have Magnus Hall effect only in TRS invariant materials with a broken inversion symmetry. This is to ensure a net non-zero Magnus contribution coming from the right movers. For example, in IS broken Dirac materials,
 the Berry curvatures have opposite signs in the two valleys due to TRS which makes it an odd function across the Brillouin zone. So, in absence of tilt or anisotropy in the dispersion, the number of right moving states at the Fermi pockets of both the valleys are equal and hence their Magnus contributions
 will cancel each other due to opposite signs of Berry curvature in the two valleys.  However, the asymmetry is not necessary in the gapped Rashba system because it has only one valley and hence an isotropic Berry curvature is not a problem. Nevertheless, the Fermi contour of the system comprises of two concentric circles for $\epsilon_F<-M$ and  $\epsilon_F>M$. For the former case, the Magnus contributions arising from the two branches of the `$-$' band have the same sign and hence no cancellation takes place. For the latter case, the Magnus contributions arise from both the bands which have opposite signs of Berry curvature. Although the signs of Magnus contributions of both the bands are opposite, they do not cancel each other because their magnitudes are different owing to the different Fermi wave vectors $k_F^+$ and $k_F^-$. Again, for $-M<\epsilon_F<M$, the Fermi contour of the system is a circle on the `$-$' band. So, there is no question of cancellation of Magnus contribution because of the same sign on Berry curvature for all the states of the `$-$' band. Thus, the Magnus effect is finite in the system despite an isotropic/symmetric Berry curvature. The Magnus contribution trivially vanishes as $M\to 0$ because the Berry curvature also vanishes in that limit.

The Magnus Hall conductivity $\sigma_m$ (in units of 
$\mathcal{\sigma}_{0}=\frac{e^2}{\hbar}\frac{\Delta U }{16\pi\epsilon_{\alpha}}$) as a function of 
scaled Fermi energy  $\tilde \epsilon_F$ for  
different values of  $\tilde{M}$ is shown in Fig. (\ref{FigMH}). Figure  (\ref{FigMH}) 
depicts two peaks at the gap edges, i.e, at $\tilde{\epsilon}_F=\pm 2\tilde{M}$, 
where the magnitudes of the peaks decrease with the increasing strength of $\tilde{M}$.
Furthermore, the Magnus Hall conductivity displays 
a plateau when Fermi energy lies between the two gap edges, i.e. 
$ -2\tilde{ M} < \tilde{\epsilon}_F <2\tilde{ M} $. For regime $(i)$ and $(iii)$,
 the Magnus Hall conductivity increases with $\tilde{M}$, which is opposite to the nature of variation of the peak values with $M$.
Since, the Berry curvature
decreases with momentum, Berry curvature for regime $(iii)$ 
is greater than that for regime $(i)$ (see Fig. \ref{Fig2}). Thus, the magnitude of peak
at $\tilde{\epsilon}_F=-2\tilde{M}$ is greater than that
at  $\tilde{\epsilon}_F= +2\tilde{M}$.  Furthermore, the peaks have opposite signs.

Unlike the systems considered in previous works \cite{Papaj,Mandal}, the gapped Rashba system has both {\it equilibrium Magnus} and {\it anomalous Hall currents} in addition to the Magnus Hall current (see Eq. (\ref{eq13})). The variation of {\it anomalous Hall current} with Fermi energy has already been studied for this system \cite{Xiao}. The {\it equilibrium Magnus current} is also expected to show similar trend of variation with Fermi energy as the {\it anomalous} one. Thus, in our work, we focus only on the Magnus Hall current. The fourth term in Eq.(\ref{eq13}) vanishes for the gapped Rashba system, due to the isotropic nature of  Fermi surface. In the gapped Rashba system, it is not possible to generate a pure Magnus Hall current.  The Magnus Hall contribution can be isolated in an experiment by simply subtracting the Hall current measured without the inbuilt potential gradient from the Hall current obtained in presence of the gradient. However, the TRS invariant systems which exhibit Magnus Hall effect will have a pure Magnus current since their {\it equilibrium Magnus} and {\it anomalous Hall} contributions will vanish. The Magnus current can be identified from their peaks when the Fermi energy is at a gap edges. The tuning of the Fermi energy can be done by applying an overall gate potential in the experimental setup.

 It is to be noted that in the diffusive regime, the Magnus Hall effect is equivalent to the nonlinear anomalous Hall effect arising due to Berry curvature dipole $(\partial_{k_x} \Omega_z)$\cite{Sodemann}. This can be explained as follows. The nonlinear anomalous Hall current is given by $j_y=\sigma_{yxx}E_x^2$ where $E_x$ is the electric field along $x$-direction and $\sigma_{yxx}\propto\int \Omega_z (\partial_{k_x} f_0)d^2{\bf k}= \int f_0 (\partial_{k_x} \Omega_z)d^2{\bf k}$ plus the boundary term. 
	On decomposing $E_x$ as $E_x=E_{in}+E_{ext}$, we get $E_x^2=E_{in}^2+2E_{in} E_{ext}+E_{ext}^2$. 
	The first term is second order in $E_{in}$, the second term corresponds to the Magnus contribution and the third term can be conventionally called as the {\it nonlinear anomalous Hall current} as it is proportional to square of the applied bias. Thus, the Magnus effect is a byproduct of nonlinear Hall effect when both internal and external fields are present. The origin of both effects are, however, the same i.e. Berry curvature dipole. In general, none of the three terms of the nonlinear Hall current can be neglected. It is only due to our intent of  highlighting a Hall effect linear in $E_{ext}$ that the $E_{in}^2$ and $E_{ext}^2$ contributions are disregarded. However, we have checked that the diffusive $\sigma_{yxx}$ vanishes for the gapped Rashba system. Hence, all the three terms of the nonlinear Hall current identically vanish for this system in the diffusive regime.
	
	The physics is slightly different in the ballistic regime. In this regime, the correction to distribution function is approximated to be independent of $E_{in}$ because of the slow variation of the inbuilt potential.  As a result, the nonlinear Hall current have contributions which go as $E_{in}E_{ext}$ and $E_{ext}^2$ only. Moreover, the nonlinear Hall current can no longer be written in terms of Berry curvature dipole by partial integration, because the form of correction to the distribution function for the ballistic case is different from the diffusive case. This can be shown as follows:	
	In the diffusive regime,
	\begin{eqnarray}
		\sigma_{yxx}\propto\iint_{k} \Omega_z (\partial_{k_x} f_0) d^2{\bf k}&=&\oint_\Gamma \Omega_z f_0 \hat{\mathbf{x}}\cdot \hat{\mathbf{n}} d\Gamma\\\nonumber&-& \iint_{k} f_0 (\partial_{k_x} \Omega_z) d^2{\bf k},
	\end{eqnarray}
	where the last time on the right hand side corresponds to Berry curvature dipole.	
In the ballistic regime,
\begin{eqnarray}
	\label{eqnlb}
	\sigma_{yxx}&\propto&\iint_{v_x>0} \Omega_z (\partial_{\epsilon}f_0) d^2{\bf k}\\\nonumber&\propto&\iint_{v_x>0} \frac{\Omega_z}{v_x}(\partial_{k_x} f_0) d^2{\bf k}=\oint_\Gamma \frac{\Omega_z}{v_x} f_0 \hat{\mathbf{x}}\cdot \hat{\mathbf{n}} d\Gamma\\\nonumber&-& \iint_{v_x>0} f_0 \partial_{k_x} (\frac{\Omega_z}{v_x}) d^2{\bf k}
\end{eqnarray}
	Clearly, the last term in Eq.(\ref{eqnlb}) is different from Berry curvature dipole. The constraint $v_x>0$ in the integrals results in finite value of $\sigma_{yxx}$ for the gapped Rashba system in the ballistic limit. In this limit, both  $E_{in}E_{ext}$ and $E_{ext}^2$ contributions survive.

Now we calculate the spin counterpart of Magnus conductivity, i.e, Magnus spin
 conductivity for all polarization directions.
 The Magnus spin conductivities in  ballistic regime are calculated 
 using Eq. (\ref{eq17}),  where we find  $\sigma_{m,yy}\neq 0$, $\sigma_{m,yz}\neq0$ and $\sigma_{m,yx}=0$.  Thus,
 the Magnus spin conductivity having polarization in $\hat{x}$
 direction, i.e, $\sigma_{m,yx}$ vanishes, which can be explained using the angular integrals: 
 	 $\sigma_{m,yx} \sim\int_{-\pi/2}^{\pi/2}~\sin\phi~d\phi~$ or $\int_{\pi/2}^{3\pi/2}~\sin\phi~d\phi$, depending on $\epsilon_F$ and $\lambda$. The $\sin\phi$ factor arises from $\sigma_x^\lambda$ i.e. $x$-component of spin polarization. Both the integrals vanish, which is clearly a mathematical fact.
 	 \begin{figure}[htbp]
 	 	\hspace{-0.4cm}\includegraphics[trim={3cm 0cm 0cm 0cm},clip,width=9.5cm]{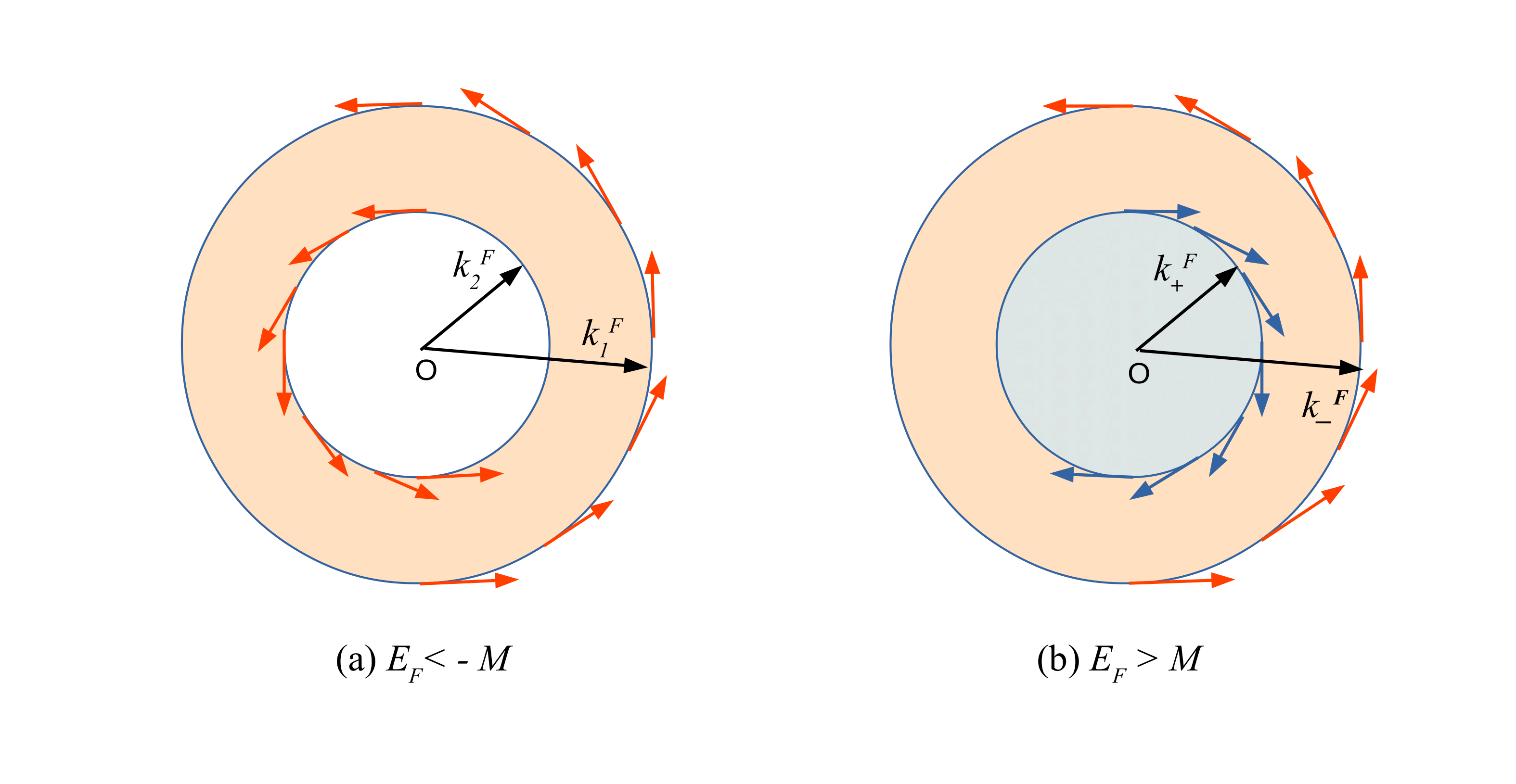}
 	 	\caption{Schematic representation of the in-plane component of spin-polarization vector (arrows) at points on the Fermi contours where $v_x>0$ for (a) $E_F<-M$ and (b) $E_F>M$. The red and blue  arrows are for $`-$' and $`+$' energy branches respectively.}
 	 	\label{figphy}
 	 \end{figure}
 	  Its physical origin can be explained as follows. From Eq. (\ref{eq17}), we see that $\sigma_{m,yx}$ is proportional to the product of $\sigma_x$  and $\Omega_z$. Since $\Omega_z$ has same sign for all the states on either of the  disjoint circles of the Fermi contour, the presence or absence of a net contribution to $\sigma_{m,yx}$ from each circle clearly depends on vector sum of $x$-component of spin-polarizations for states with $v_x>0$.
  In the 2D Rashba system discussed in our manuscript, the spin-polarization vector is always perpendicular to the velocity of the electrons i.e. $\langle \hat{\boldsymbol{\sigma}}^\lambda \rangle\cdot {\bf v}^\lambda=0$ (see Fig.\ref{figphy}). Due to this arrangement, the $x$-component of spin polarizations on upper and lower halves of the contour cancel pairwise  when summed over all states with $v_x>0$. This accounts for the vanishing of $\sigma_{m,yx}$. However, this is not a universal feature. Had the Rashba interaction been of the form $\boldsymbol{\sigma}\cdot{\bf k}$, the spin-velocity locking would be of the form $\langle \hat{\boldsymbol{\sigma}}^\lambda \rangle \times {\bf v}^\lambda=0$. In that case, the in-plane component of spin polarization vector will be parallel to the velocity. Under that arrangement, the Magnus spin conductivities with polarization along the bias would be finite and that perpendicular (in-plane) to it would vanish.
\begin{figure}
\begin{center}
\includegraphics[width=83mm,height=54mm]{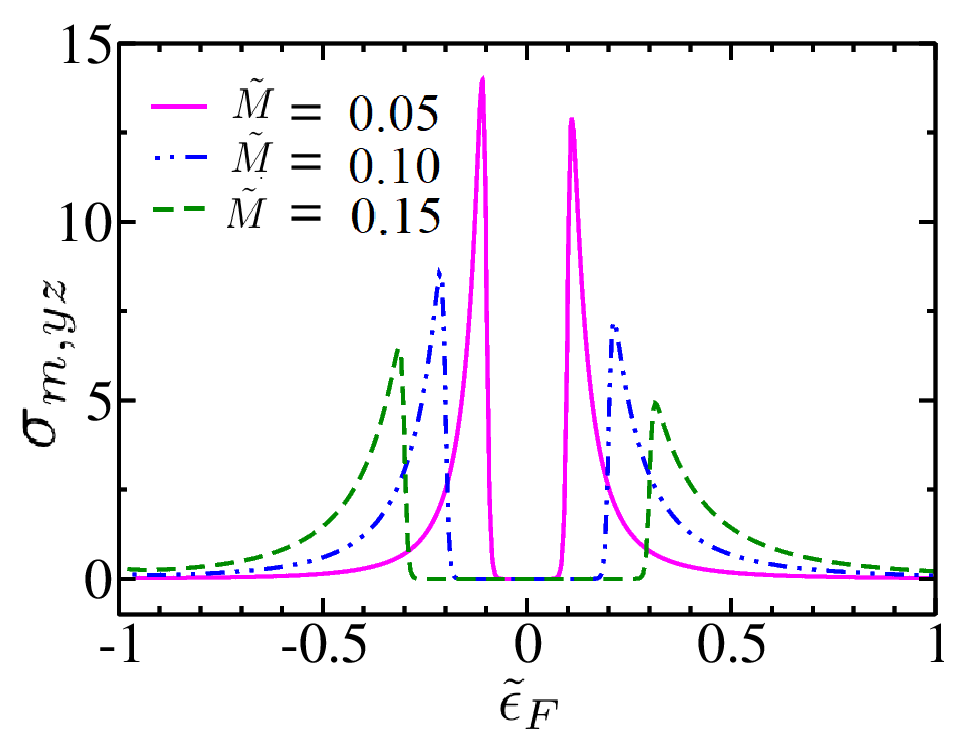}\caption {The Magnus spin Hall conductivity  $\sigma_{m,yz}$ 
(in units of $\sigma_{1}$)
 as a function of scaled Fermi energy  $\tilde{\epsilon}_F$  for three different values of $\tilde{M}$. Temperature is fixed at $T=1$ K.}
\label{FigMS1}
\end{center}
\end{figure}  

At zero temperature the Magnus spin conductivity having polarization
 in $\hat{z}$ direction for ballistic regime is obtained as
\begin{eqnarray}
\sigma_{m,yz} &=&\sigma_{1}\frac{2\tilde{M}^2(\tilde{E}_F^2+3)}{(\tilde{\epsilon}_F+\tilde{M}^2)^3}
\hspace{0.03in} \mathrm{for \hspace{0.03in} regime} \hspace{0.03in} (i),  \\\nonumber
\sigma_{m,yz} &=&\sigma_{1}\frac{\tilde{M}^2}{\tilde{E}_F(\tilde{E}_F+1)^3}\hspace{0.1in} \mathrm{for \hspace{0.03in} regime} \hspace{0.05in} (ii),\\\nonumber 
\sigma_{m,yz} &=&-\sigma_{1}\frac{2\tilde{M}^2(3\tilde{E}_F^2+1)}{\tilde{E}_F(\tilde{\epsilon}_F+\tilde{M}^2)^3}
\hspace{0.04in} \mathrm{for \hspace{0.03in} regime} \hspace{0.03in} (iii),
\end{eqnarray}
where $\sigma_{1}=\frac{e}{32\pi}\frac{\Delta U}{\epsilon_{\alpha}}$.
For $\sigma_{m,yz}$, the applied electric field, the propagation and polarization
 directions are mutually perpendicular to each other and hence it can be called as Magnus spin Hall conductivity.

\begin{figure}
\begin{center}
\includegraphics[width=80mm,height=60mm]{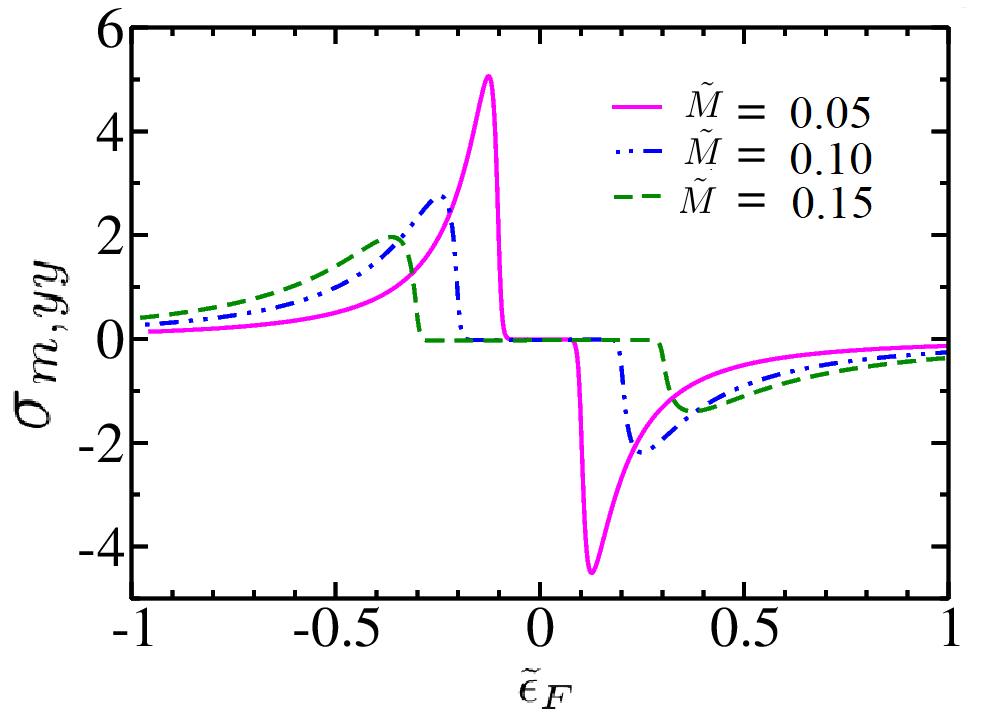}
\caption {The Magnus spin conductivity  $\sigma_{m,yy}$ (in units of $\sigma_{1}$)
 as a function of $\tilde{\epsilon}_F$  for different values of $\tilde{M}$. Temperature is fixed at $T=1$ K.}
\label{FigMS2}
\end{center}
\end{figure}  

The Magnus spin Hall conductivity  $\sigma_{m,yz}$ (in units of $\sigma_{1}=\frac{e}{32\pi}\frac{\Delta U}{\epsilon_{\alpha}} $) as
 a function of $\tilde{\epsilon}_F$  for three different values of $\tilde{M}$ is shown in Fig. \ref{FigMS1}. Figure \ref{FigMS1}
 reveals that the Magnus spin Hall conductivity also has the peaks at the gap edges which have  same signs unlike the Magnus Hall conductivity.
 Similar to $\sigma_m$, the magnitudes of the peaks decrease with $\tilde{M}$, but for regimes $(i)$
 and $(iii)$, the magnitude of $\sigma_{m,yz}$ increases with $M$.

 The gapped Rashba system has both spin
Hall and Magnus spin Hall contributions. Both spin Hall and Magnus spin Hall effects arise on application of a source-drain bias (external field).
But, to realize the latter, an inbuilt electric field is also required in addition to the bias. A finite Berry curvature is
necessary for Magnus spin Hall effect, but not for spin Hall effect.
Unlike the spin
 Hall conductivity which has a universal value $\sigma_s=\frac{e}{8\pi}$ for $M\rightarrow0$ \cite{Sinova1,Kapri}, the Magnus 
spin Hall conductivity vanishes for $M\rightarrow0$.
 This occurs due to the fact that the definition of spin Hall conductivity
 is weighted by zeroth order distribution function ($f_0$), whereas
 Magnus spin Hall conductivity is weighted by first order distribution function ($f_1$). In case of spin Hall 
conductivity, we need to perform the integration, where the upper limit yields
 $-\frac{e}{16\pi} [\frac{\tilde{M}^2}{(\tilde{E}_F-1)^2}+\frac{M^2}{(\tilde{E}_F+1)^2}]$ (which vanishes for $M\rightarrow0$) and the lower limit
 of integration yields the universal value $\frac{e}{8\pi}$. On the contrary, for Magnus
 spin Hall conductivity, the integration yields the value of integrand at Fermi energy  
(because to the presence of  first order distribution function
 ($f_1\propto\frac{\partial f_0}{\partial \epsilon}=-\delta(\epsilon-\epsilon_F)$) 
which vanishes at $M\rightarrow0$ (see Eq. (\ref{eq17}) and Eq. (\ref{eq23}) ).

The Magnus spin conductivity $\sigma_{m,yy}$  having polarization in $\hat{}y$ 
direction at zero temperature for ballistic regime  is obtained as
\begin{widetext}
\begin{eqnarray}
\sigma_{m,yy} &=&-\frac{2\sigma_{1}}{\pi}\frac{\tilde{M}}{\tilde{E}_F}\Big[\frac{(\tilde{E}_F+1)^3[(\tilde{E}_F-1)^2-\tilde{M}^2]^{1/2}+(\tilde{E}_F-1)^3[(\tilde{E}_F+1)^2-\tilde{M}^2]^{1/2}}{(\tilde{\epsilon}_F+\tilde{M}^2)^3}\Big]\hspace{0.03in} \mathrm{for \hspace{0.03in} regime} \hspace{0.03in} (i) \mathrm{\hspace{0.03in} and \hspace{0.05in}} (iii), \\\nonumber
\sigma_{m,yy} &=&-\frac{2\sigma_{1}}{\pi}\frac{\tilde{M}}{\tilde{E}_F}\Big[\frac{[(\tilde{E}_F+1)^2-\tilde{M}^2]^{1/2}}{(\tilde{E}_F+1)^3}\Big]\hspace{0.1in} \mathrm{for \hspace{0.05in} regime} \hspace{0.05in} (ii).
\end{eqnarray}
\end{widetext}

Figure \ref{FigMS2} shows the Magnus spin conductivity  $\sigma_{m,yy}$
 (in units of $\sigma_{1}$)
 as a function of $\tilde{\epsilon}_F$  for different values of $\tilde{M}$.
 As expected,  $\sigma_{m,yy}$ also shows the peaks at at $\tilde{\epsilon}_F=\pm 2\tilde{M}$.
  Unlike the case of $\sigma_{m,yz}$,  the peaks have opposite signs. As earlier,
  the peak magnitude decreases with $\tilde{M}$ and for the regime $(i)$ and $(iii)$, the reverse nature is obtained.

\subsection{Thermally driven Magnus conductivity and Magnus spin conductivities}

\begin{figure}
\begin{center}
\includegraphics[width=82mm,height=60mm]{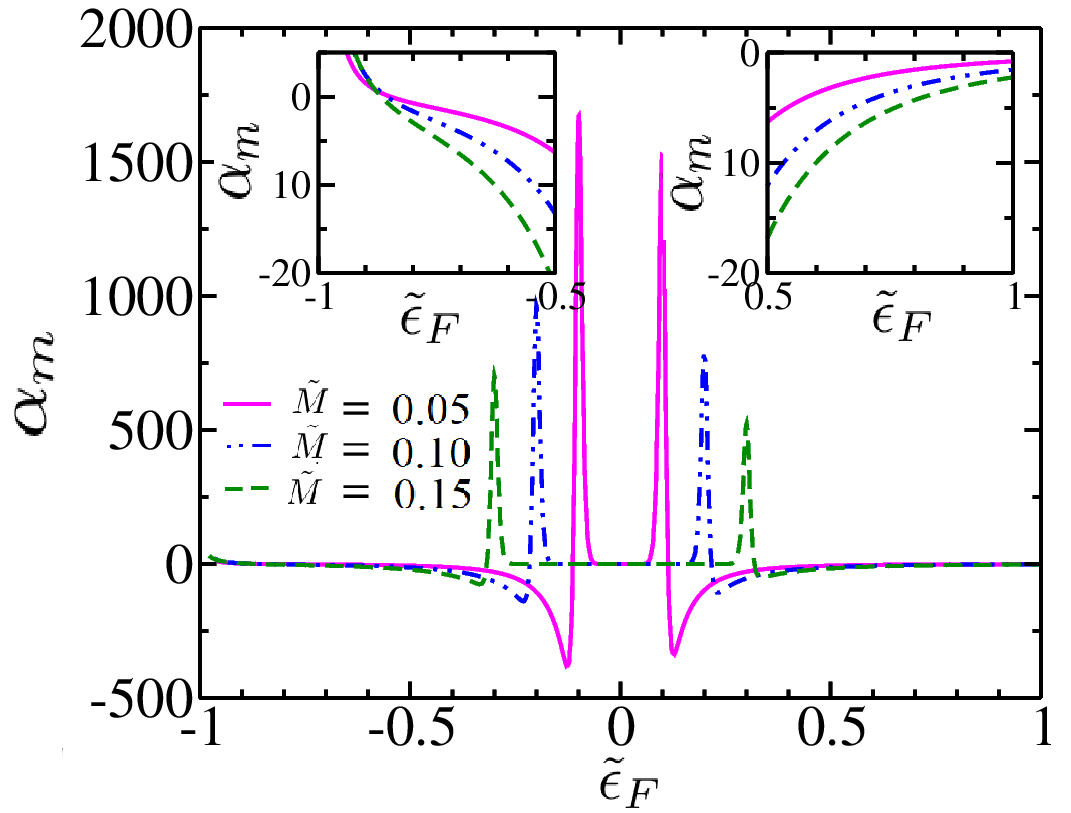}
\caption {The Magnus Nernst conductivity $\alpha_m$ (in units of 
$\mathcal{\alpha}_{0}$)
 as a function of 
scaled Fermi energy  $\tilde \epsilon_F$ for  
different values of  $\tilde{M}$. Temperature is fixed at $T=1$ K.}
\label{FigN}
\end{center}
\end{figure}

In previous section, we have studied electric field ($\hat{x}$ directed) driven Magnus conductivities, while this
 section presents  the results of  thermally driven (because of a temperature gradient
($\nabla_x T $) between the source and drain) Magnus  conductivities .

First we calculate the Magnus Nernst conductivity (thermal analogue of Magnus Hall conductivity) using Eq. (\ref{eq12}).
  In the limit $\epsilon_F\gg k_B T $, the Magnus Nernst conductivity for ballistic regime has the following forms
\begin{eqnarray}
\label{eq26}
\alpha_m &=-&\mathcal{\alpha}_{0}\frac{16\tilde{M}}{(\tilde{\epsilon}_F+\tilde{M}^2)^3}\hspace{0.03in} \mathrm{for \hspace{0.03in} regime} \hspace{0.03in} (i)  \\\nonumber
\alpha_m &=&\mathcal{\alpha}_{0}\frac{\tilde{M}(1+3\tilde{E}_F)}{\tilde{E}_F^3(1+\tilde{E}_F)^3}\hspace{0.1in} \mathrm{for \hspace{0.05in} regime} \hspace{0.05in} (ii),\\\nonumber
\alpha_m &=&\mathcal{\alpha}_{0}\frac{\tilde{M}(6\tilde{E}_F^4+12\tilde{E}_F^2-2)}{\tilde{E}_F^3(\tilde{\epsilon}_F+\tilde{M}^2)^3}\hspace{0.05in} \mathrm{for \hspace{0.03in} regime} \hspace{0.03in} (iii), 
\end{eqnarray}
where $\mathcal{\alpha}_{0}=\frac{e\Delta U\pi k_B^2T}{96\hbar\epsilon_{\alpha}^2}$. Here, it is to be noted that the above
expressions are valid for the Fermi energies away from the gap edges.

The Magnus Nernst conductivity $\alpha_m$ (in units of 
$\mathcal{\alpha}_{0}=\frac{e\Delta U\pi k_B^2T}{96\hbar\epsilon_{\alpha}^2}$) as a function of 
scaled Fermi energy  $\tilde \epsilon_F$ for  
different values of  $\tilde{M}$ is shown in Fig. (\ref{FigN}). Unlike the previous cases, the peaks
at the gap edges are accompanied by kinks, 
where the peak and kink values decrease with the increasing strength of $\tilde{M}$. The kinks in Fig. \ref{FigN} 
can not be captured by the analytical expressions in Eq. (\ref{eq26}).  For each value of $M$, the plot displays nearly a plateau when Fermi energy lies inside the gap.
 For regime $(i)$ (see right panel of insets) and $(iii)$ (see left panel of insets), the Magnus
 Nernst conductivity increases with $\tilde{M}$.
Unlike the Magnus Hall conductivity, the peaks for Magnus Nernst conductivity
 have  same signs because of the term $\epsilon^{\prime}_k=(\epsilon_k-\epsilon_F)$  in Eq. (\ref{eq12}).

Now we present the results of thermally driven Magnus spin conductivities with different polarization in ballistic limit. 
 Similar to the case of electric field driven Magnus spin conductivities, here also
 we find that Magnus spin conductivities having polarization
 in $\hat{z}$ and $\hat{y}$ directions are non zero, i.e. $\alpha_{m,yy}\neq 0$,
 $\alpha_{m,yz}\neq0$, whereas the Magnus spin conductivity
 having polarization in $\hat{x}$-direction ($\alpha_{m,yx}$) vanishes because of the $\phi$ integration. The physical reason is same as earlier.

\begin{figure}
\begin{center}
\includegraphics[width=80mm,height=60mm]{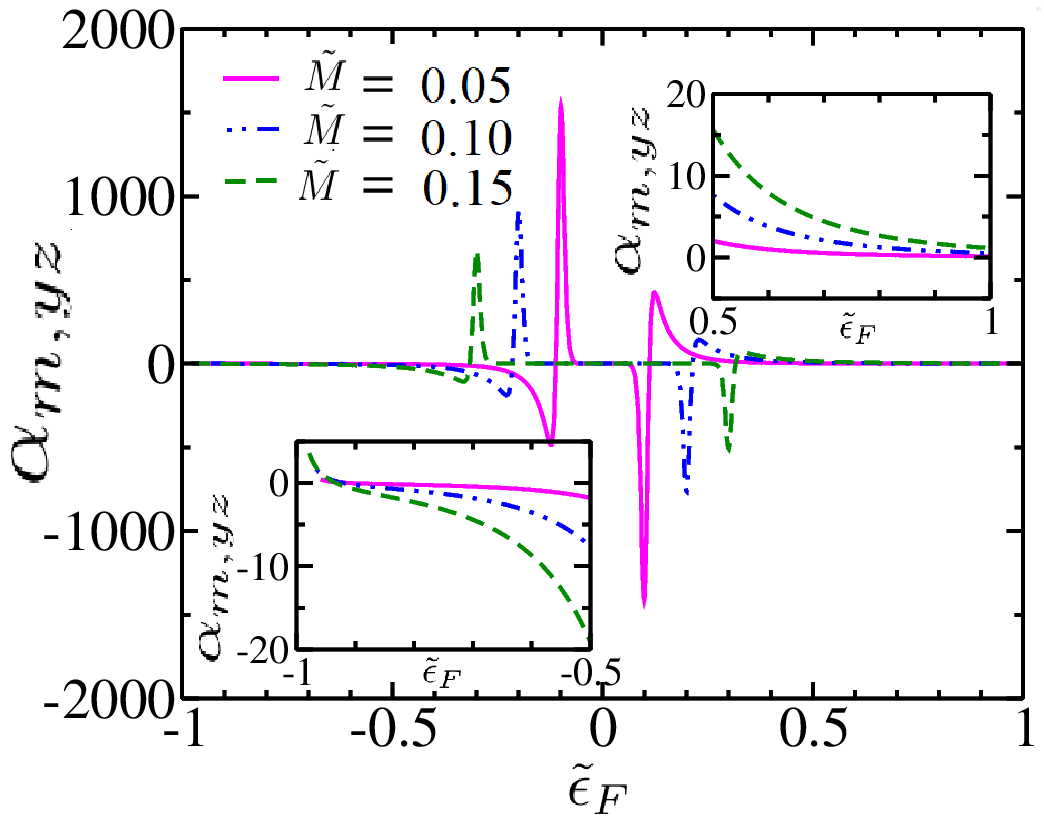}
\caption {The Magnus spin Nernst conductivity  $\alpha_{m,yz}$
 (in units of $\alpha_{1}$)
 as a function of $\tilde{\epsilon}_F$  for different values of $\tilde{M}$. Temperature is fixed at $T=1$ K.}
\label{FigNS1}
\end{center}
\end{figure}
In the limit $\epsilon_F\gg k_B T$, the thermally driven (for $\hat{x}$ directed
 temperature gradient) Magnus spin conductivity with polarization in $\hat{z}$
 direction and propagation in $\hat{y}$ direction for ballistic regime is obtained as
\begin{eqnarray}
\alpha_{m,yz} &=& \alpha_{1}\frac{8\tilde{M}^2(\tilde{E}_F^2+5)}{(\tilde{\epsilon}_F+\tilde{M}^2)^4} \hspace{0.03in} \mathrm{for \hspace{0.03in} regime} \hspace{0.03in} (i), \\\nonumber
\alpha_{m,yz} &=& \alpha_{1}\frac{\tilde{M}^2(4\tilde{E}_F+1)}{\tilde{E}_F^3(\tilde{E}_F+1)^4}\hspace{0.1in} \mathrm{for \hspace{0.05in} regime} \hspace{0.05in} (ii),\\\nonumber
\alpha_{m,yz} &=&  -\alpha_{1}\frac{2\tilde{M}^2(15\tilde{E}_F^4+10\tilde{E}_F^2-1)}{\tilde{E}_F^3(\tilde{\epsilon}_F+\tilde{M}^2)^4} \hspace{0.03in} \mathrm{for \hspace{0.05in} regime} \hspace{0.03in} (iii),
\end{eqnarray}
where  $\alpha_{1}=\frac{\Delta U\pi k_{B}^2 T}{192\epsilon_{\alpha}^2} $.
As earlier, these expressions are valid for the Fermi energies away from gap edges.
 This can be viewed as the thermal analogue of Magnus spin Hall conductivity, 
and hence called as Magnus spin Nernst conductivity. 

Figure (\ref{FigNS1}) depicts the Magnus spin Nernst conductivity 
 $\alpha_{m,yz}$ (in units of $\alpha_{1}=\frac{\Delta U\pi k_{B}^2 T}{192\epsilon_{\alpha}^2} $)
 as a function of $\tilde{\epsilon}_F$  for different values of $\tilde{M}$,
 where the peaks and kinks near the gap edges have opposite signs, unlike the cases of Magnus spin Hall
 conductivity $\sigma_{m,yz}$ and  Magnus Nernst conductivity  $\alpha_m$.
 As expected, the behavior of $\alpha_{m,yz}$ as a function of $\tilde{M}$
 is similar as earlier, i.e, the peak and kink values decrease with increasing 
strength of $\tilde{M}$ and the opposite trend is obtained for the regime $(i)$ (see right panel of insets) and $(iii)$ (see left panel of insets).

Now, in the limit $\epsilon_F\gg k_B T$ and for the Fermi energies away from the gap edges, the  thermally driven
 Magnus spin conductivity with polarization and propagation in $\hat{y}$
 direction for ballistic regime has the following forms
\begin{widetext}
	\begin{eqnarray}
		\alpha_{m,yy} &=&\frac{2\alpha_{1}}{\pi}\frac{\tilde{M}}{\tilde{E}_F^3}\sum_{\eta=\pm}\frac{(-3{\tilde{E}_F}^3+\eta7{\tilde{E}_F}^2-5\tilde{E}_F
			+4\tilde{M}^2\tilde{E}_F+\eta-\eta\tilde{M}^2)}{[(\tilde{E}_F-\eta)^2-\tilde{M}^2]^{1/2}(\tilde{E}_F-\eta)^4}
	\hspace{0.02in} \mathrm{for \hspace{0.02in} regimes} \hspace{0.02in} (i) \mathrm{\hspace{0.02in} and \hspace{0.02in}} (iii), \\\nonumber
	\alpha_{m,yy} &=&\frac{2\alpha_{1}}{\pi}\frac{\tilde{M}}{\tilde{E}_F^3}\frac{(-3{\tilde{E}_F}^3-7{\tilde{E}_F}^2-5\tilde{E}_F
		+4\tilde{M}^2\tilde{E}_F-1+\tilde{M}^2)}{[(\tilde{E}_F+1)^2-\tilde{M}^2]^{1/2}(\tilde{E}_F+1)^4}\hspace{0.1in} \mathrm{for \hspace{0.05in} regime} \hspace{0.05in} (ii).
\end{eqnarray}
\end{widetext}

\begin{figure}
\begin{center}
\includegraphics[width=89mm,height=68mm]{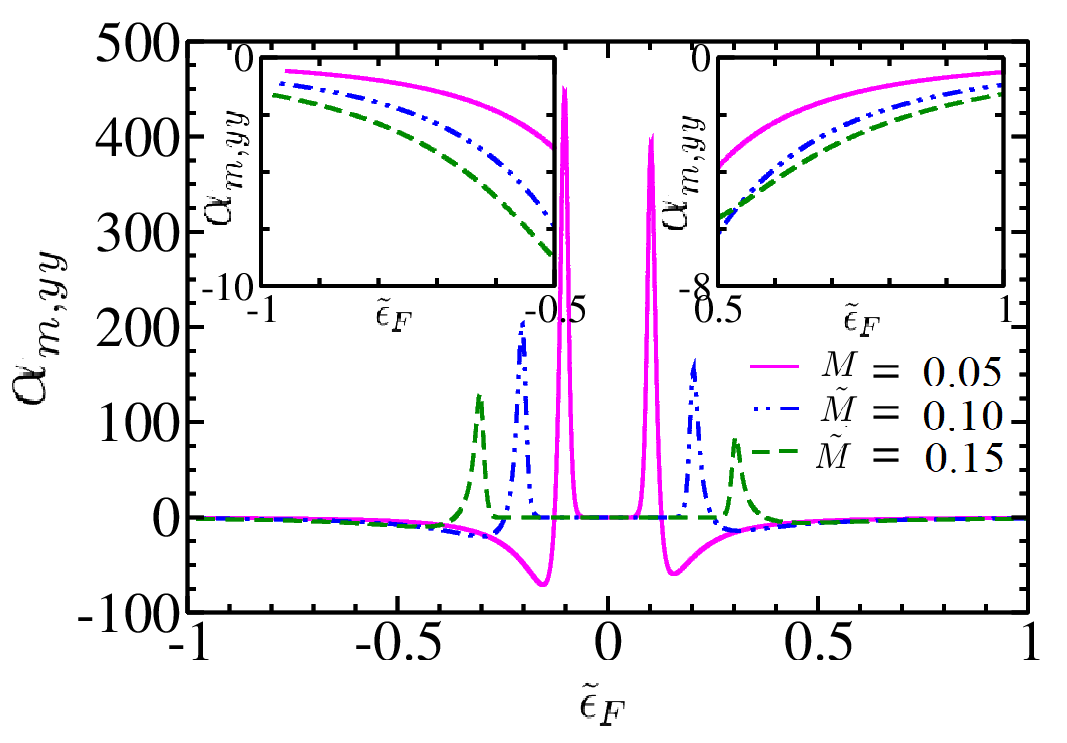}
\caption {The thermally driven Magnus spin conductivity  $\alpha_{m,yy}$ (in units of $\alpha_{1}$)
	as a function of $\tilde{\epsilon}_F$  for different values of $\tilde{M}$. Temperature is fixed at $T=1$ K.}
\label{FigNS2}
\end{center}
\end{figure}

The thermally driven Magnus spin conductivity  $\alpha_{m,yy}$ (in units of $\alpha_{1}$) as a function of $\tilde{\epsilon}_F$ 
 for different values of $\tilde{M}$ is shown in Fig. \ref{FigNS2}.
 Figure \ref{FigNS2} reveals that  $\alpha_{m,yy}$ also has the 
peaks and kinks at around the gap edges with the peaks and kinks having same signs. Insets of Fig. \ref{FigNS2}
depict the behavior of $\alpha_{m,yy}$ in the regime $(i)$ and $(iii)$, where it is clear that the magnitude of $\alpha_{m,yy}$ increases with $\tilde{M}$, though the changes are small.

 It is to be noted that all the results presented in our manuscript are for ballistic regime only ($v_x \tau\gg L$). We have checked that the minimum value of $v_x \tau$ (where $\tau$ is energy dependent) for $\phi=0$, $\alpha=0.1$ eV nm, $m^*=0.3~m_e$ and $\tau_0=2.5$ ps is $\sim 20$ nm. So, our calculations for ballistic regime are valid when the system length ($L$) is few nanometers.

 There are some qualitative differences between Magnus spin Hall and Nernst effects in the diffusive (i.e. in presence of impurities) and the ballistic regime. In the ballistic regime, only the carriers having $v_x > 0$ on the Fermi contour contribute while in the diffusive regime, all the states on the Fermi contour contribute. The diffusive conductivities have an additional factor $v_x \tau$ in the integrand. In the ballistic limit, $\sigma_{m,yz},\sigma_{m,yy}\neq0$ and $\sigma_{m,yx}=0$. 
	However, in the diffusive regime, 
	 we find that $\sigma_{m,yz}$ ($\propto\int_{0}^{2\pi}~\cos \phi~d\phi=0$) does not survive,  $\sigma_{m,yx}$ ($\propto\int_{0}^{2\pi}~\cos \phi \sin\phi~d\phi=0$) still remains zero and $\sigma_{m,yy}$ ($\propto\int_{0}^{2\pi}~\cos^2 \phi~d\phi\neq 0$) gets renormalized by a factor in regimes (i) and (ii) but has a modified expression in regime (iii). The reason is as follows: for regime (iii), the contributions of $\nu=2$ branch have opposite signs for ballistic ($\sigma_{m,yy}\propto\int_{\pi/2}^{3\pi/2}~\cos \phi~d\phi$) and diffusive regimes  ($\sigma_{m,yy}\propto\int_{0}^{2\pi}~\cos^2 \phi~d\phi$).  We have also checked that $\sigma_{m,yy}$  with $k$-dependent $\tau$  shows qualitatively similar trend of variation with Fermi energy as that with constant $\tau$. 
	 Thus, in presence of impurities, only the Magnus spin conductivities with an in-plane polarization transverse to the bias is present for the gapped Rashba system. Similar behavior is obtained by the Magnus Nernst conductivities.
 
  As mentioned earlier, our results are presented with the conventional definition of spin current operator. With the modified definition of spin current operator of Ref. \cite{Shi},  we find that the additional term in spin current operator ($\hat{r}(d{\hat s}_z/dt)$) has no contribution to $\sigma_{m,yz}$, because the torque density trivially vanishes for all states i.e. $
 \mathcal{T}_z({\bf r})=(1/i\hbar)Re \Psi^{\dag}({\bf r})[\hat{\sigma}_z,\hat H]\Psi({\bf r})=0$ .  Similarly, we find that $\mathcal{T}_y({\bf r})=(1/i\hbar)Re \Psi^{\dag}({\bf r})[\hat{\sigma}_y,\hat H]\Psi({\bf r})=\mathcal{T}_x({\bf r})=(1/i\hbar)Re \Psi^{\dag}({\bf r})[\hat{\sigma}_x,\hat H]\Psi({\bf r}) =0$.  Thus, $\sigma_{m,yj}$ presented in our manuscript remain same with the new definition of spin current operator. 
 It is worthwhile to mention that many works on spin current \cite{Sonin,Hamamoto,Pan} have been carried out with the conventional definition of spin current operator.
\section{Conclusion}
\label{sec5}
In this paper, we have explored the Magnus transport
in a system, where both the inversion symmetry and
 time reversal symmetry are broken with a finite Berry curvature.
The Magnus Hall effect arises because of the Magnus velocity which appears in a self-rotating
quantum electronic wave-packet moving through a crystalline material having potential
energy gradient. We have studied both the electric
 field driven and temperature gradient driven Magnus conductivities 
in a  gapped 2D Rashba system in ballistic regime. The spin counterparts
 of the Magnus conductivities have also been explored by modifying
the spin current operator with inclusion of the Magnus velocity. Unlike the spin
 Hall conductivity which has a universal value $\sigma_s=\frac{e}{8\pi}$ as $M\rightarrow0$, the Magnus 
spin Hall conductivity vanishes in this limit. We have found that  the
Magnus spin currents with spin polarization perpendicular to the applied bias (electrical/thermal) are finite while with polarization along the bias vanishes because of the $\phi$ integration.
We have studied the role of gap parameter ($M$, induced by the time 
reversal symmetry breaking term) and the Fermi energy ($\epsilon_{F}$) on 
the behavior of different  conductivities. We find that all the conductivities 
and their spin counterparts show peaks (and kinks for thermally driven conductivities) at the gap edges, where 
magnitudes of peaks (and kinks) decrease with the increasing strength of gap parameter.
 For the regime 
$\epsilon_{F}> M$ and $\epsilon_{F}<- M$, the conductivities increases with increasing strength of $M$. All the Magnus conductivities are nearly constant when the Fermi energy varies inside the gap.  

\section{Acknowledgment
}
P. Kapri thanks Department of Physics, IIT Kanpur,
India and Department of Physics, Osaka University, Japan.

\end{document}